\documentstyle[10pt]{article}

\def\bfl{\begin{flushleft}}
\def\efl{\end{flushleft}}
\def\bfr{\begin{flushright}}
\def\efr{\end{flushright}}
\def\bc{\begin{center}}
\def\ec{\end{center}}
\def\be{\begin{equation}}
\def\ee{\end{equation}}
\def\ba{\begin{eqnarray}}
\def\ea{\end{eqnarray}}
\def\nn{\nonumber }

\def\text#1{\mbox{#1} }

\def\drm{d}

\def\Sign#1{\, \text{sign}\left(#1\right) }
\def\Lim#1#2{\, \lim\limits_{#1 \to #2} }

\begin{document}

~\\
~\\
~\\
\bfr
gr-qc/9802042\\
Mod. Phys. Lett. A 13 (1998) 1419-1425\\
\efr
~\\
~\\
~\\
\bc
{\LARGE \bf
Mass of perfect fluid black shells
}

~~\\
{\large Konstantin G. Zloshchastiev}\\
~~\\
Department of Theoretical Physics,
Dnepropetrovsk State University,\\
Nauchniy lane 13, Dnepropetrovsk 320625,
Ukraine.\footnote{E-mail address: zlosh@usa.net}\\

\ec

~~\\

\abstract{
The spherically symmetric singular perfect fluid shells are considered 
when their radii are equal to the event horizon (the black shells).
We study their observable masses, depending at least on the three 
parameters, viz. the square speed of sound in the shell, 
instantaneous radial velocity of the shell at a moment 
when it reaches the horizon, 
and integration constant related to surface mass density.
We discuss the features of black shells depending on an equation 
of state.
}

~\\



\large
\newpage

Since the classical works \cite{dau,isr} there has been a lot of progress
on the investigation of 
thin shells in general relativity
(see Refs. \cite{sat,mae,bkt}).
It was found that shells are both simple and instructive models of 
several dynamical and cosmological objects and processes.
The formalism of the thin shell theory has been widely described  in the 
literature
(see Ref. \cite{mtw} for details), therefore, we shall only point out 
the most important features.
In this letter we study a class of spherically symmetric shells 
of some perfect fluids.
Attention will be paid to their behaviour near the event horizon.

One considers the matter thin layer with the surface stress-energy 
tensor of a perfect fluid in the general case 
(we use the units $\gamma=c=1$, where 
$\gamma$ is the gravitational constant)
\[
S_{ab}=\sigma u_a u_b + p (u_a u_b +~ ^{(3)}\!g_{ab}),
\]
where $\sigma$ and $p$ are the surface mass-energy density  and  pressure 
respectively, {\bf u} is the time-like unit tangent vector, 
$~^{(3)}g_{ab}$ is the three-metric on a shell.
We suppose the metrics of the space-times outside $\Sigma^+$ and inside 
$\Sigma^-$ of a spherically symmetric shell to be of the form
\be
\drm s_\pm^2 =
-[1+\Phi^{\pm}(r)] \drm t^2_\pm + [1+\Phi^{\pm}(r)]^{-1} \drm 
r^2 + r^2 \drm \Omega^2,                                       \label{eq1}
\ee
where $d\Omega^2$ is the metric of the unit two-sphere.
It is possible to show that if one introduces the proper time  
$\tau$, then the 3-metric of a shell can be written in the form
\be
^{(3)}\!\drm s^2 = - \drm \tau^2 + R^2 \drm \Omega^2,        \label{eq2}
\ee
where $R(\tau)$ is a proper radius of a shell.
Define a simple jump of the second fundamental forms across a shell as 
$[K^a_b]=K^{a+}_b - K^{a-}_b$, where 
\[
K^{a\pm}_b = 
            \lim\limits_{n\to \pm0} 
            \frac{1}{2} ~^{(3)}\!g^{a c} 
            \frac{\partial}{\partial n} ~^{(3)}\!g_{c b},
\]
where $n$ is a proper distance (time-like or space-like 
in the general case, space-like in our case) in the normal direction.
The Einstein equations on a shell then read \cite{bkt,vis}
\ba
&&\sigma = -\frac{1}{4\pi} [K^\theta_\theta],                \label{eq3}\\
&&p = \frac{1}{8\pi} ([K^\tau_\tau] + [K^\theta_\theta]).     \label{eq4}
\ea
Besides, an integrability condition of the Einstein equations is
the energy conservation law for shell matter.
In terms of the proper time it can be written as
\be
\drm \left( \sigma ~^{(3)}\!g \right) +
p~ \drm \left( ~^{(3)}\!g \right) 
+ ~^{(3)}\!g~   [ T ]\, \drm \tau =0,                        \label{eq5}
\ee
where $[ T ] = (T^\tau_n)^+ - (T^\tau_n)^-$,  
$T^\tau_n = T_\alpha^\beta u^\alpha n_\beta$ is the
projection of stress-energy tensors in the $\Sigma^\pm$
space-times on the tangent and normal vectors, 
$^{(3)}\!g=\sqrt{-\det{(^{(3)}\!g_{ab})}} = R^2 \sin{\theta}$.
In this equation, the first term corresponds to a change in
the shell's
internal energy, the second term corresponds to the work done
by the internal forces of the shell, while the third term
corresponds to the flux of energy across a shell.

For clarity we study below the class of the black shells with the 
most simple event horizons,
therefore, we suppose the space-times (\ref{eq1}) to be the Schwarzschild's 
(thereby the special case of the Minkowski flat space-time is also 
included).
We assume
\be
\Phi^\pm = - \frac{~2 M_\pm}{R},                              \label{eq6}
\ee
where $M_\pm$ is the total energy of the configuration with respect to 
static distant observers in the space-times $\Sigma^+$ and $\Sigma^-$ 
respectively.
It corresponds to a body with the mass $M_-$ surrounded by a shell.
Then a computation of the extrinsic curvatures yields
\ba
&&R K^{\theta \pm}_\theta =  \epsilon_\pm 
\sqrt{1 + \dot R^2 - \frac{2 M_\pm}{R}},                      \label{eq7}\\
&&K^{\tau \pm}_\tau = \frac{1}{\dot R}
\frac{\drm}{\drm \tau} (R K^{\theta \pm}_\theta),             \label{eq8}
\ea
and we can write Eq. (\ref{eq3}) in the form
\be
\epsilon_+ \sqrt{1+\dot R^2 - \frac{2 M_+}{R}} - \epsilon_-
\sqrt{1+\dot R^2 - \frac{2 M_-}{R}} = - \frac{m}{R},        \label{eq9}
\ee
where
\be
m = 4 \pi \sigma R^2                                 \label{eq10}
\ee
is interpreted as the effective rest mass, 
$\dot R=\drm R/\drm\tau$ is a proper velocity of a
shell, $\epsilon_\pm = \Sign{\sqrt{1+\dot R^2 - 2 M_\pm/R }}$.

It is well-known that $\epsilon = +1$ if $R$ increases in the outward
normal direction to the shell (e.g., it takes place in a flat 
space-time),
and $\epsilon = -1$ if $R$ decreases (semiclosed world).
Thus, only under the additional choice $\epsilon_+ = \epsilon_-=1$  we 
have an ordinary (black hole type) shell  \cite{bkt,bkkt,gk}.
It seems to be the most physical case.
For definiteness in the letter we deal only with such shells, the 
rest of the cases can easily be considered by analogy.

As for the conservation law (\ref{eq5}), one can obtain that
$[ T ]$ is identically zero for the Schwarzschild space-times (\ref{eq6}).
Further, we assume the equation of state of the shell matter to be
that of some perfect fluid, viz.,
\be
p=\eta \sigma.                                         \label{eq11}
\ee
This equation includes the most studied cases: the dust shell $p=0$
\cite{isr,bkkt,hb},
radiation fluid shell $\sigma - 2 p=0$ \cite{vis}, and bubble 
$\sigma + p =0$ \cite{bkt,cl,lm}.
If $\eta >0$, it can be interpreted as a square component of the vector of a 
speed of sound in the shell.
Then for a spatially two-dimensional homogeneous fluid the
square speed of sound is $2 \eta$.
From the physical viewpoint some $\eta$ appear to be inadmissible.
For instance, if a fluid is required to satisfy the dominant energy 
condition, $\sigma \geq |p|$, one obtains the constraint
\be
|\eta| \leq 1.                                           \label{eq12}
\ee
If a fluid is required to satisfy the causality condition,
we get the constraint
\be
\eta \leq 1/2,                                            \label{eq13}
\ee 
where one takes into account spatial two-dimensionality of a fluid.
Nevertheless, the aim of this letter is to study the general case of 
arbitrary $\eta$.
So, solving the differential equation (\ref{eq5}) with respect to $\sigma$,
we obtain
\be
\sigma  = \frac{\sqrt{2C}}{4\pi} R^{-2 (\eta + 1)},         \label{eq14}
\ee
where $C$ is the integration constant determined by the specific shell's
matter.
The value of $C$ is closely related to the value of surface mass density
(or pressure) at fixed $R$.
We consider ordinary shells not wormholes \cite{bkt,vis}, therefore, 
$\sigma \geq 0$ is required.
It should also be noted that from Eq.\ (\ref{eq9}) at positive 
densities $\sigma$,
it follows that $M_+ > M_-$ for any $R$, $\dot R$ and $m(R) \not = 0$.
Otherwise, matching of the space-times is impossible.
The equations (\ref{eq9}), (\ref{eq10}), and (\ref{eq14}) together  
with the choice of the signs $\epsilon_\pm$ 
completely determine the motion of the perfect fluid shell.
So, we have all the necessary equations to consider the 
behavior of the shells near the horizon.

First of all we consider what an external observer will see when the
shell collapses.
The Lichnerowicz-Darmois-Israel formalism gives us the relation between 
the time of a static observer in the external space-time $\Sigma^+$
and proper time of the shell.                                     
Indeed, taking into account of (\ref{eq6}), 
the gauge of the four-velocity yields
\[
-\left( 1 - \frac{2 M_+}{R} \right) 
\dot t_+^2 + \frac{\dot R^2}{1 - \frac{2 M_+}{R}} = -1,
\]
and thus, the shell radial velocity with respect to a static external 
observer is determined by the expression
\be
\left( \frac{\drm R}{\drm t_+} \right)^2 = 
\frac{\left( 1 - \frac{2 M_+}{R} \right)^2}
     {1+\dot R^2 - \frac{2 M_+}{R}} \dot R^2.               \label{eq15}
\ee
It can easily be seen that this velocity asymptotically vanishes at the
horizon point $R=2 M_+$,  
therefore, under external distant observation the black shell is static.
Thus, we can also consider the black shell as a model of a black hole.
What are the parameters of the  objects obtained?
What is the difference between the black shells made of several fluids?
These are problems we must resolve.

So, let us consider the fluid shell passing through the external event 
horizon $R=2 M_+$.
Define the instantaneous radial velocity of the shell at a moment, when 
it reaches the horizon, as
\[
v = \dot R |_{R=2 M_+}.
\]
We assume $v \not = 0$, otherwise $K^\tau_\tau$ becomes infinite (\ref{eq8}).
Then the equation of motion (\ref{eq9})  yields the equation with respect 
to the external mass $M_+$ of a black shell in the general case
\be
2 M_+ \sqrt{1+ v^2 - \frac{M_-}{M_+}} = m |_{R=2 M_+}.        \label{eq16}
\ee
For our black shells, taking into account Eqs. (\ref{eq10}) and
(\ref{eq14}), we have
\be
\sqrt{
     1+v^2 - \frac{M_-}{M_+}  
     } = \sqrt{2 C}\, (2 M_+)^{-2\eta-1}.              \label{eq17}
\ee
We obtain the algebraic equation with respect to $M_+$.
Thus, the dependency of the external (observable) total mass-energy on the 
internal one is nonlinear and the total mass-energy $M_+$ is not a sum 
$M_- + f(C,\eta)$, as it would take place in the non-relativistic case.
This is a simple reflection of the known fact, that within 
the framework of general relativity the energy superposition principle is 
not valid as a rule.
Also it can readily be seen that when the black shell is formed, 
the kinetic energy is converted into the observable total mass too. 

Let us perform the analysis of the obtained equation for several $\eta$.
It comes easy to us because we can consider the inverse function 
$ M_- (C, \eta, v; M_+)$.
Naturally we divide all shells into the five classes with respect to $\eta$.
These cases are illustrated in Fig.\ref{fig1}.
The physical sector seems to be defined by the non-negative 
$M_+$, $M_-$ masses (though a theory of quantum gravity could give 
rise to space-times of negative mass, see Ref. \cite{man}).
Hence, in the figure the physical sector lies on the right of the $M_+$ axis.

(a) $\eta > -1/4$.
This is the most physically admissible class, because it includes the class
$\eta > 0$.
It is easy to see, that $\Lim{M_+}{0} M_- = -\infty$ and 
$\Lim{M_+}{+\infty} M_- = (1+ v^2) M_+$, i.e., we have the behavior,
qualitatively described by the curve {\it a}, see Fig.\ref{fig1}.
It should be pointed out that $M_+ \not = 0$ at $M_- = 0$, i.e., the
shells have a proper gravitational mass and can exist in absence of the 
``stuffing'' $M_-$ (the so called hollow black shells).
For instance, the proper Schwarzschild radii of the dust and 
radiation fluid shells are respectively
\[
2 M_+= \sqrt{ \frac{2 C}{1+v^2}},~~
2 M_+= \sqrt[4]{ \frac{2 C}{1+v^2}}.
\]

(b) $\eta = -1/4$.
In this case the function (\ref{eq17}) is simply the line 
({\it b}, Fig.\ref{fig1}) $M_- = M_+ (1+v^2) - C$.  
The proper gravitational mass of this black shell is the observable mass
of the corresponding hollow black shell, $M_+ = C/(1+v^2)$.

(c) $-1/2 < \eta < -1/4$.
In this case the curve $M_- (M_+)$ has the single minimum point
\be
M_{-\text{min}} = -2 (2\eta + 1) C^{\frac{1}{4\eta + 2}}
                  \left[
                        -\frac{1+ v^2}{2 (4 \eta +1)}
                  \right]^{\frac{4 \eta +1}{4\eta + 2}},    \label{eq18}
\ee
at 
\be
M_{+\text{min}} = \frac{1}{2} 
                  \left[
                        -\frac{2 C (4 \eta +1)}{1+v^2}
                  \right]^{\frac{1}{4\eta + 2}}.        \label{eq19}
\ee
One can see from Fig.\ref{fig1} (the curve {\it c}) that $M_- = 0$
at the two values of $M_+$ 
\be
M_+ = 0,~
M_+ = \frac{1}{2}
      \left(
             \frac{2 C}{ 1+v^2}
      \right)^{\frac{1}{4\eta + 2}}.                      \label{eq20}
\ee
Thus, for this class of black shells we have the ambiguity in determining
the mass.
Of course, this ambiguity  takes place only for hollow black shells, because
for stuffed black shells it lies in the unphysical sectors 
$M_- < 0$ or $M_+ <0$.

({\it d}) $\eta = -1/2$.
In this case we also have a line 
({\it d}, Fig.\ref{fig1}), $M_- = M_+ ( 1+v^2 - 2 C)$.  
It is of interest to note that the proper gravitational mass of 
such black shells is zero, i. e. they
can be observable only at $M_- \not = 0$.
In other words, if the space-time inside such shells is flat,
then the external space-time will also be flat.
It can easily be explained by the fact that the matter with 
the equation of state $\sigma + 2 p =0$ has to be a two-dimensional 
analog of the three-dimensional global texture 
$^{(3)}\!\varepsilon + 3 ~^{(3)}\!p =0$,
which is the topological defect having zero gravitational mass \cite{dad}.

({\it e}) $\eta < -1/2$.
This case, including the (black) bubble at $\eta = - 1$, is a mirror 
reflection of the case ({\it c}).
The curve $M_- (M_+)$ has the single maximum point given by the expressions
(\ref{eq18}), (\ref{eq19}), just replace the subscript ``min'' by ``max''.
Here we also have the ambiguity in determining the mass 
(except the maximum point (\ref{eq18}), (\ref{eq19}) 
and $M_- = 0$ point (\ref{eq20})), 
but now it lies in the physical sector $M_- > 0$, $M_+ > 0$ 
(see curve {\it e}).
Therefore, these black shells can additionally be divided into the two 
subtypes with respect to the two-valued mass.
Finally we calculate the proper total mass of the black bubble.
From Eq. (\ref{eq17}) one obtains
\be
M_+ = \frac{1}{2} \sqrt{\frac{1+v^2}{2 C}},
\ee
where the second root $M_+ = M_- = 0$ was rejected as trivial.

Thus, in this letter the classification of some barotropic
perfect fluid black thin shells was considered by means of the standard 
Lichnerowicz-Darmois-Israel formalism, 
thereby the nonlinear behaviour of their total mass was of special interest.

\section*{Appendix: Frequently Asked Question}

When this paper was published some people suggested me
that Eq. (16) was wrongly ruled out from Eq. (9).
It seems to be true for a first look.
Indeed, when we blindly substitute $R=2 M_+$ into (9) we must obtain
\[
2 M_+ 
\left(
\sqrt{1+ v^2 - \frac{M_-}{M_+}} - v
\right) = m |_{R=2 M_+}       
\]
instead of (16).
Let us show where the mistake appears.
Eq. (9) follows from the equation
\[
(K_{\theta\theta})^+ -  (K_{\theta\theta})^- = - m,
\]
which when calculating extrinsic curvature $K_{\theta\theta}$
can be written as
\be                                                 \label{eq22}
(n^r)^+ - (n^r)^- = - m/R,
\ee
where $n^{r\pm}$ are radial components of normal vector.
Each of them is obtained from the equations \cite{isr}:
\ba
&&g_{00} (u^t)^2 - g_{00}^{-1} (u^r)^2 = -1, \nn\\
&&u^t n_t + u^r n_r = 0, \nn\\
&&g_{00} (n^t)^2 - g_{00}^{-1} (n^r)^2 = 1, \nn
\ea
where $u^r=\dot R$, and $g_{00} =-(1-2M/R)$ for our case.
Thus we have 3 equations for 3 unknown variables $u^t, n^t, n^r$.
We find that
\be           \label{eq23}
n^r = \pm \sqrt{g_{00} + \dot R^2},
\ee
and Eq. (9) is evident.
However, at $R=2 M$ we have $g_{00}=0$ 
whereas last expression was ruled out in the assumption 
$R \not= 2 M$.
In other words, $g_{00}=0$ is the singular point of the system above
and should be considered separately.
We obtain
\[
\frac{(n^{r+})^2|_{R=2 M^+}}{0} = 1,
\]
therefore, $(n^r)^+ = 0$ at the horizon point $R=2 M^+$.
It does not seem to be extraordinary: the surface of black hole is known to 
be null, and vectors {\bf n} and {\bf u} are 
degenerated.
Thus, the first term in (\ref{eq22}) 
vanishes whereas the second is obtained
from (\ref{eq23}), i.e. Eq. (16) is true.

Therefore, physical picture seems to be as follows.
The hypersurfaces $R=\mbox{const}$ 
have to be timelike when $R > R_{R=2M_+}$, 
null at $R = R_{R=2M_+}$,
and spacelike at $R < R_{R=2M_+}$.
Therefore, if we wish to preserve the definition of the shell as 
the surface $R=R (s)$ (where $s$ 
is some evolution parameter) then we cannot rigidly fix
whether a (two-sided) shell has timelike, spacelike or null surfaces 
for all $R$ (see also 
V. de la Cruz and W. Israel, Nuov. Cim. A 51 (1967) 744).
Only then we obtain a model of the true collapse of standard 3D matter 
when the continuous set of timelike matter layers forms 
the black hole having null surface.

\def\CJP  {Czech. J. Phys. }
\def\CQG  {Class. Quantum Grav.}
\def\EPL  {Europhys. Lett.}
\def\GRG  {Gen. Rel. Grav.}
\def\NPh  {Nucl. Phys.}
\def\PhE  {Phys. Essays}
\def\PhL  {Phys. Lett. }
\def\PhR  {Phys. Rev.}
\def\PhRL {Phys. Rev. Lett.}
\def\PhRp {Phys. Rep.}
\def\NCim {Nuovo Cim.}
\def\NuPh {Nucl. Phys.}

\def\Name#1{#1,}
\def\Review#1{#1}
\def\Book#1{{\sl #1}}
\def\Vol#1{#1}
\def\Year#1{(#1)}
\def\Page#1{#1}
\def\And{and }

\def\jn#1#2#3#4#5{{#1}{#2} {#3} {(#5)} {#4}}
\def\boo#1#2#3#4#5{{\it #1} ({#2}, {#3}, {#4}){#5}}
\def\prpr#1#2#3#4#5{{``#1,''} {#2}{#3}{#4}, {#5} (unpublished)}


\begin{figure}
\vbox to 1cm{\vfill\centerline{
\fbox{The figure}}\vfill}
\caption{The qualitative plot of the observable total mass 
$M_+$ versus internal one for several $\eta$: 
({\it a}) $\eta > -1/4$, 
({\it b}) $\eta = -1/4$, 
({\it c}) $-1/2 < \eta < -1/4$,
({\it d}) $\eta = -1/2$,
({\it e}) $\eta < -1/2$.}
\label{fig1}
\end{figure}


\newpage

\begin{thebibliography}{99}

\bibitem{dau}
G. Dautcourt, \jn{Math. Nachr.}{}{27}{277}{1964}.

\bibitem{isr}
\Name{W. Israel} 
\Review{Nuovo Cim. B} \Vol{44} \Year{1966} \Page{1}.

\bibitem{sat}
\Name{H. Sato} 
\Review{Prog. Theor. Phys.} \Vol{76} \Year{1986} \Page{1250}.

\bibitem{mae}
\Name{K. Maeda} 
\Review{\GRG} \Vol{18} \Year{1986} \Page{931}.

\bibitem{bkt}
\Name{V. A. Berezin, V. A. Kuzmin, \And I. I. Tkachev} 
\Review{Phys. Rev. D} \Vol{36} \Year{1987} \Page{2919}.

\bibitem{mtw}
\Name{C. W. Misner, K. S. Thorne, \And J. A. Wheeler}
\Book{Gravitation} (Freeman, San Francisco) \Year{1973}.
                                              

\bibitem{vis}
\Name{M. Visser} 
\Review{Nucl. Phys. B} \Vol{328} \Year{1989} \Page{203}.

\bibitem{bkkt}
\Name{V. A. Berezin, N. G. Kozimirov, V. A. Kuzmin, \And I. I. Tkachev} 
\Review{\PhL B} \Vol{212} \Year{1988} \Page{415}.

\bibitem{gk}
\Name{D. S. Goldwirth \And J. Katz} 
\Review{\CQG} \Vol{12} \Year{1995} \Page{769}.

\bibitem{hb}
P. H\'aj\'\i\v{c}ek and J. Bi\v{c}\'{a}k, 
\jn{\PhR}{ D}{56}{4706}{1997};
J. L. Friedman, J. Louko, and S. N. Winters-Hilt, 
\jn{{\it ibid.}}{}{56}{7674}{1997}.


\bibitem{cl}
S. Coleman and F. De Luccia, \jn{Phys. Rev.}{ D}{21}{3305}{1980}.

\bibitem{lm}
P. Laguna-Castillo and R. A. Matzner, \jn{Phys. Rev.}{ D}{34}{2913}{1986};
B. Jensen, \jn{Phys. Rev.}{ D}{51}{5511}{1995}.

\bibitem{man}
R. Mann, \jn{\CQG}{}{14}{2927}{1997}.

\bibitem{dad}
D. Notzold, \jn{Phys. Rev.}{ D}{43}{R961}{1991};
N. Dadhich, Preprint No. IUCAA-60/97 (1997).

\end{thebibliography}
\end{document}